# Understanding the Effect of Agile Practice Quality on Software Product Quality

Sherlock A. Licorish
School of Computing
University of Otago
Dunedin, New Zealand
sherlock.licorish@otago.ac.nz

*Abstract*—Agile methods and associated practices have been held to deliver value to software developers and customers. Research studies have reported team productivity and software quality benefits. While such insights are helpful for understanding how agile methods add value during software development, there is need for understanding the intersection of useful practices and outcomes over project duration. This study addresses this opportunity and conducted an observation study of student projects that was complemented by the analysis of demographics data and open responses about the challenges encountered during the use of agile practices. Data from 22 student teams comprising 85 responses were analyzed using quantitative and qualitative approaches, where among our findings we observed that the use of good coding practices and quality management techniques were positively correlated with all dimensions of product quality (e.g., functionality scope and software packaging). Outcomes also reveal that software product quality was predicted by requirements scoping, team planning and communication, and coding practice. However, high levels of team planning and communication were not necessary for all software development activities. When examining project challenges, it was observed that lack of technical skills and poor time management present most challenges to project success. While these challenges may be mitigated by agile practices, such practices may themselves create unease, requiring balance during project implementation.

## I. Introduction

Agile methods, processes and practices are held to assist software developers in many ways [1-3], often leading to good project outcomes [4]. Thus, the use of agile methods has become mainstream [2, 5]. Agile methods encourage incremental development in an environment that promotes continuous improvement. The processes here are the recommended guidelines, while the practices refer to the actual implementation of the guidelines. In agile software development methodologies such as Scrum, Kanban, Extreme Programming (XP) and Feature-Driven Development (FDD), there is a gradual surfacing of the software design and requirements, which promotes a humanistic environment, having persons interacting closely, employing a 'speculate-collaborate-learn' approach. This approach to software development is often termed 'lightweight'. Agile practices are typically implemented in contrast to the previous waterfall life cycle, which is held to be documentation driven, in a continuous 'plan-build-implement framework' [6].

Studies have reported several benefits related to the use of agile methods and practices, including productivity gains [7], improvement in software quality [8, 9], improvement in team morale [10] and team empowerment [11]. That said, while useful insights are provided by these works, most of these studies tend to employ interviews and surveys, reflecting largely on practitioners' opinions. This is because researchers often face challenges associated with accessing professional software engineering teams to study their actual practices and project data, beyond open-source projects [12-14]. Thus, there is need to analyze the full breath of the agile software development lifecycle as well as consider multiple success dimensions of the products that are produced by agile developers to further understand the utility of agile methods. Student projects have contributed to this cause, in aiding our understanding around how successful agile teams perform [15, 16] and a range of other issues [17, 18]. However, there remain gaps in our knowledge around the specific contexts for when agile practices hinder versus improve teams. Our work contributes to this body of knowledge.

The contributions of this paper are as follows:
- We present an approach for studying software practices and product quality through a range of student project data, providing methodological pointers for other researchers.
- We provide evidence into the way software development practice and product quality are related, and the specific practices that predict software quality – this offers a granular view into the use of agile practices over the software development life cycle and their impact on product outcomes.
- Our evidence is useful for practitioners using agile methods and will support the appropriate use of specific techniques by practitioners. For instance, teams should commit to good requirements engineering practice, as this helps during software design and coding. A commitment to good coding practices (including automated testing) will lead to high quality software releases.
- We advance the state of research into agile methods and

their utility, contributing to the body of knowledge (research) on the appropriateness of agile methods and practices and provide recommendations for future research.

The remaining sections of this manuscript are organized as follows. We review the literature and develop our research questions in the next section (Section II). We next present the study settings and methodology in Section III, before providing the results in Section IV. Section V discusses our findings and their implications. We then consider the threats to the work in Section VI, concluding the study with a summary and future research directions in Section VII.

## II. Literature Review and Research Questions

**Benefits Associated with Agile Methods**: In terms of the benefits that Agile methods deliver, Ahmed et al. [7] analyzed agile methodologies to assess the benefits associated with their use in organizations through a survey, where among 42 responses practitioners noted that the most used methodology was Scrum, and 67% of the practitioners mentioned that their productivity and software quality improved through the use of this approach. Tarhan and Yilmaz's [4] study discovered that teams using agile processes performed better than those using regular incremental process in terms of productivity, software defect density, defect resolution effort ratio, test execution effectiveness and effort prediction capability. Wadood et al. [8] surveyed public sector in-house software development projects and highlighted the benefits of using agile methodologies for software quality in the public sector. Additionally, their findings demonstrated that agile software development approaches significantly affect the quality of software products and successful software delivery within budget and deadline. Zainab et al. [10] surveyed 24 companies on their use of agile approaches where it was found that agile teams had high morale and the project sponsors were involved. Also, these teams typically had optimal testing regime and benefited from limiting work in progress.

Jamieson's [19] position paper emphasized several benefits to using agile methods, including increased flexibility, improved maintenance capabilities, decreased development time, and other efficiencies. Jain et al. [20] established that agile software development process attributes may be mapped to the five phases of the software development life cycle, where this mapping may help with simplicity. These works have all provided support for agile methods use, albeit the context and specific usage scenario may also need consideration. For instance, Poth et al. [21] emphasized that mature agile teams actively request external "feedback" which supports improvement efforts, meaning that those that are less matured may not reap the benefits associated with regular client input which is typical of agile methods use. That said, excessive client involvement may lead to technical debt [1]. In addition, Lindsjørn et al. [22] survey involving responses from 477 respondents from 71 teams in 26 companies showed lower agreement among the raters regarding the evaluation of agile team performance than was the case for traditional teams surveyed.

On the contrary, Rico's [9] survey aimed at measuring the relationships between the four factors of agile methods and scholarly models of websites' quality showed that the use of iterative development and customer feedback are linked to higher website quality. In addition, agile methods were linked to improvements in cost efficiency, productivity, quality, cycle time, and customer satisfaction. Iqbal et al.'s [11] survey of agile teams' productivity found that apart from leader meetings and unit and regression testing, which were negatively correlated with productivity, several other factors affected productivity positively regardless of contexts. Of note were team empowerment, inter-team coordination, requirements as user stories, requirements workshop involvement, clear features, test cases and integration testing. Abdalhamid et al. [23] explored the quality that is gained by adopting agile methods where pair programming and test-driven development (TDD) were noted to be the best choices for increasing software quality.

**Augmenting Agile Methods**: Beyond the benefits that are derived from agile methods' use, the proposition of augmenting agile methods with other processes and frameworks for improvements is supported by multiple studies. Veeranjaneyulu et al. [27] argue for quality assurance to be intertwined in agile process as against being seen as a separate activity after development if teams are to deliver high quality software products. Shahzeydi et al. [28] explored continuous delivery as part of agile projects, noting that the quality of continuous delivery directly impresses the quality of the development process. These authors noted that there is need for efficiencies in people, organization, processes and tools for enhanced software quality. Heck and Zaidman [29] examined relevant quality criteria for assessing the correctness of agile requirements through a literature review, where requirements relevance over project duration was established as essential in agile contexts. The idea here is that teams are encouraged to only document what is relevant at a given moment and postpone all other requirements documentation to as late as possible in the software life cycle. Almeida et al. [30] investigated the benefits of the combined adoption of DevOps and agile, finding that benefits related to automation, communication, and time to market are noteworthy when these approaches are combined.

**Knowledge Gaps and Research Questions**: Results above seem to support the position that enhanced software development practice and product quality may be achieved by following recommended agile processes (e.g., [4]). Also, although agile methods are studied widely [31], and there have been many attempts to verify the effectiveness of agile methodologies (e.g., [2, 4, 7, 8, 10, 19, 32]), this empirical evidence is not complete. Furthermore, most of the evidence tend to be self-reported by practitioners, largely through the conduct of surveys and interviews, due to scarcity in access to professional software engineering teams to study their actual practices and project data. In fact, gains associated with the use of agile methods are not evident in all circumstances, as others have shown that context and team maturity are also important factors (e.g., [21]), as are team dynamics and other factors [15]. Developers have also reported various other challenges when adopting agile approaches [5]. While these works provide useful insights into agile methods, there is need to analyze the full breath of the software development lifecycle as well as consider multiple success dimensions of the products that are produced to further investigate the utility of agile methods, in terms of when specific agile practices add value.

Student projects have helped our understanding in this regard [15, 16], given the flexibility afforded with such projects (i.e., as noted above, real software development teams operating in industry rarely provide their software project data for research, beyond open-source projects). Notwithstanding the limitations in student settings (refer to Section VI for further details), they provide value if environments are close to what ensues in industry, as is the case for the current study.

Specific practices tend to be the basis of many agile approaches, as embedded in such teams' way of working. For instance, most agile methods embrace requirements change and incremental development and delivery. Through actual observation of real project data, even those performed by students, we would be afforded answers to questions such as: Do these practices aid or hinder teams? Does agile software development practices quality align with the quality of software development outcomes? Which agile software development practices impact on the quality of software development outcomes? We explore these issues in this work, providing several contributions to the body of knowledge and practice by answering the following research questions.

**RQ1**. How do software practice and product performances vary among agile software development teams?

**RQ2**. What are the relationships between variations in the use of software development practices and software product performance?

**RQ3**. Are there particular team challenges that affect software practice and product performances during agile software development?

### III. STUDY SETTINGS

*A. Data Collection*

This study followed up on a web-based survey which was aligned with the IEEE and ACM 2014 Software Engineering Education Knowledge (SEEK) and curriculum [16, 38], aimed at understanding students' opinion of their project performance in relation to the adequacy of the outcomes they developed. The IEEE and ACM SEEK and curriculum provides guidance on what software engineering graduates should know and how such skills should be taught if graduates are to lead a successful software engineering career. The guide includes professional knowledge (e.g., about standards and ethics), technical knowledge (e.g., covering problem analysis and design, development, testing, and deployment), teamwork (e.g., in relation to self-motivation, time management, team communication and teamwork), end-user awareness (e.g., user experience, client negotiation, leadership and questioning skills), design solutions in context (e.g., understand different design guidelines for various domains), perform trade-offs (e.g., involving feasibility evaluations, refining goals and objectives and ability to compromise), and foster continuing professional development (e.g., curiosity and initiative, ability to learn and thirst for knowledge and critical thinking). University of Otago's software engineering curricula is designed to ensure the aforementioned knowledge areas are taught in preparing professional software engineers.

As part of data collection, we followed previous surveys [39, 40] in recording students' age, gender, academic major, year of study and the job the student aspire to do in the future. We then captured students' previous experience, including if the students worked in a project team before, were employed in a salaried job, previously learned about software engineering, previously occupied a team role, previously acquired technical skills, the average number of hours committed weekly by the students, the name of the project they were involved and their project team size (all demographics data). We also collected open-ended responses to triangulate our quantitative data, capturing insights around the biggest challenges students faced during their software development projects. The survey[1] received ethical approval and some of the outcomes were published [16]. We would be happy to share administrative access to the instrument with researchers interested in running the survey. In the current work, we collect additional data, analyze data previously collected which were not analyzed and accompany this survey data with further observations we collected from assessment data in relation to the software development practice and product quality to satisfy the current study objectives.

Undergraduate students at University of Otago that completed a software engineering course where a major software project was done were invited to participate in the survey in 2019 (for the earlier study [16]) and 2020 (new data). The course is lectured over 13 weeks, and is taken in year three in the Computer Science or Information Science/Software Engineering majors (Bachelors level). At the end of the software engineering course, students would have practised the fundamental skills required to develop software systems using modern tools, practices and development environments (see Table I). Software engineering courses at University of Otago build on, apply and extend material introduced in previous courses (e.g., software engineering processes, analysis, design, programming skills, programming paradigms, testing, web and mobile development, human computer interaction, etc). These courses are practice-based, and they provide the first opportunity for students to undertake a sizeable piece of practical work that spans sufficient time to expose the complexities of modern software development, mimicking a semi-professional environment.

Students undertaking the software engineering course typically operate in teams of between 4 to 6 members, and frequently participate in industrial projects involving an external client. Projects were balanced based on the number of members present in the team. Previous research has shown that companies use these projects for: recruiting developers, getting software developed and piloting/researching new technologies [41]. Student projects traversed processes of requirements elicitation and scoping (done via interviews), negotiation of schedules, coding, verification and testing (see Table I). The central point of assessment for the courses is the project itself rather than a test or exam. Assessments cover business vision, use cases, requirements and stakeholder analysis, prototypes, design documentation, risk analysis, project plans, the actual software product, user documentation, issue tracking, source code, technical documentation, testing and test protocols (unit, acceptance, integration, etc.), and deployment (see Table I). To obtain individual grades, instructors also consider source code repositories, issue trackers, project management logs, contribution statements and activity logs, and experiences from close collaboration with the students.

---
[1] https://tinyurl.com/y6nfqfl7

TABLE I. SOFTWARE DEVELOPMENT TECHNIQUES, TOOLS, PROGRAMMING COURSES, ARTEFACTS AND TEAM ASSIGNMENT METHOD

| Techniques Used | Tools Used | Avg Programming Courses Completed | # Artefacts Provided | Team Assignment Method |
|---|---|---|---|---|
| Up front analysis to understand domain and develop initial (but evolving/iterative) business idea, incremental, expert estimation, iterative development, weekly stand-ups, weekly retrospections, coaching and feedback sessions with staff, continuous integration, and unit testing. Teams followed Scrum. | Git (GitBucket and GitHub), Gradle, Taiga, Slack, Discord, Trello, NetBeans, Visual Studio, JUnit, Eclipse, Android Studio. | 6 | User stories (requirements), product backlog, estimation and schedule details, team assignment details, design documents, prototypes, risk register, project plans, issue tracker, the actual product, user documentation, source code, technical documentation, unit tests, and acceptance tests. | Educators' and Students' assigned (students could request to work with or avoid specific members). |

Technical tutorials and mentoring sessions further validate students' individual contributions. As noted in Table I, students implement their projects using (Scrum) agile practices, involving incremental development with some up-front analysis to understand the domain and decompose initial (but evolving) business ideas. Students practised weekly stand-ups, weekly retrospections, other team meetings, coaching and feedback sessions with staff, and weekly technical tutorials on tools and technologies (e.g., build systems and configuration management, version control, unit testing and continuous integration). Students also aim for frequent delivery and incorporate continuous client feedback. As part of software delivery, students are expected to employ good software packaging and deployment techniques, and they use various software metrics for product and process improvement. The students studied completed six programming courses on average prior to the software engineering course. The artefacts provided by students covered the entire software development life cycle, from project scoping and requirements documents to the tested and packaged software. Team assignment was supervised by the university staff to ensure diversity and fairness. Students could also request to work with or avoid specific members, where these requests were granted.

As noted above, data (demographics and open comments in relation to project challenges) were collected over 2019 and 2020. In 2019, 79 students were enrolled in the software engineering course (71 local and 8 international), where 58 agreed to participate in the study (one response was removed as the student did not clearly identify their project, leaving 57). For 2020, 41 students were enrolled (35 local and 6 international), where 28 students agreed to participate in the study. Given the disparities in numbers for local and international students in the courses we did not differentiate between the two groups during data collection and analysis. Also, all students in the courses were completing the local university degree for the past three (3) years, so there was no difference in the actual education that these students were receiving.

Altogether, 86 students participated in the study, however, as noted above one response was removed as the student did not clearly identify their project teams, leaving 85 useful responses (60 males, 25 females) and project observations that were suitable for analysis. This represents a 70.8% response rate (see Section IV for further details). For most projects, more than half of the members in the team participated in the study, thus providing a detailed view of students' project realities. Table II provides a summary of the projects, where it is seen that projects (22 altogether) covered mostly web and mobile applications. Languages include Java, JavaScript, C#, Ruby, HTML, CSS, SQL, .NET, PHP and Python.

TABLE II. PROJECTS SUMMARY

| Project Name* | Description | #Members Studied | Team Size |
|---|---|---|---|
| System 1 | Web application for childcare facility | 3 | 5 |
| System 2 | Web application for managing students collaboration | 5 | 5 |
| System 3 | Management System Web application for contract and customer relationship management | 2 | 5 |
| System 4 | Mobile application to digitise student ID services | 5 | 5 |
| System 5 | Web application for online shopping | 4 | 5 |
| System 6 | Mobile application for managing events | 2 | 5 |
| System 7 | Mobile application to manage recipes and caloric details | 3 | 5 |
| System 8 | Tourism mobile application for events and festivals | 4 | 6 |
| System 9 | Web application for managing elderly companionship | 5 | 6 |
| System 10 | Web application for managing an image library | 6 | 6 |
| System 11 | Web application for managing library occupancy | 4 | 5 |
| System 12 | Mobile application for managing medical prescriptions | 5 | 5 |
| System 13 | Mobile App Mobile application for supporting international students | 4 | 6 |
| System 14 | Mobile application for managing pet database | 3 | 6 |

| Project Name* | Description | #Members Studied | Team Size |
|---|---|---|---|
| System 15 | Mobile application to manage receipt | 3 | 4 |
| System 16 | Web application for performing business analytics | 1 | 5 |
| System 17 | Web application that manages restaurant bookings | 5 | 5 |
| System 18 | Mobile application for managing students' attendance | 4 | 5 |
| System 19 | Web application for timetable management | 4 | 5 |
| System 20 | Web application for managing student marketplace | 5 | 5 |
| System 21 | Mobile application for managing printing services | 5 | 6 |
| System 22 | Mobile application for managing groceries and pantry | 3 | 6 |

Note: * Project names adjusted to maintain anonymity

*B. Constructs and Measures*

As noted above, the software teams studied used Scrum, and were rewarded for the degree of conformance to recommended processes (**guidelines**), see below. The software engineering body of knowledge identify these software development processes (e.g., requirements scoping) around the analysis, design, construct, test pipeline [42], and the most established software models emphasized that the quality of software product should cover functionality, reliability, usability, efficiency, maintainability and portability [43]. Of note here is that these latter dimensions not only consider the software features and their appropriate scope and working order (e.g., functionality and usability), but also the appropriateness of the software to facilitate future maintenance and easy portability. Accordingly, various constructs were independently assessed (out of 100) throughout the project for software development practices implemented across five dimensions below. Practices (**process implementations**) were evaluated over the 13 weeks project duration, with weekly and continuous observations and evaluations, contributing to a final score/measure. The quality of the final product was evaluated along four dimensions out of 100 (see below), where detailed evaluations were performed after the course ended and the projects were finally released. The final project evaluations at the end of the course were done independent of the practices' evaluations during the course (the full rubric is available here[2]). In terms of weighting, the product was graded for twice as much as the process. Teams were assessed as a whole where all members contributed equally, however, at times individuals were assessed separately where it was evident that their contributions stood out remarkably (less or more than the team's). Teams' practice and product outcomes were used for our investigation, as follows.

**1) Practice Dimensions (measured via evaluators' scores):**
- *Requirements Scoping*–Agile (Scrum) requirements engineering processes are properly adopted, covering elicitation, specification, validation and management. Requirements are iteratively captured and exhaustive, covering the full list of features, services and constraints. Requirements are also properly expressed as user stories and changes are facilitated over the project duration, where the backlog is used as recommended in the agile body of knowledge.
- *Estimation and Scheduling*–Planning poker (expert estimation) is used properly where estimates are realistic and accurate. Scheduling of user stories in sprints is meaningful, taking into account feature dependencies, user story size, team members' ability and the team's velocity. The product backlog, release backlog and sprint backlog are used appropriately, and the team has a shared vision around the use of these repositories and artefacts.
- *Team Planning and Communication*–Team planning covers collaborative, iterative, continuous work on the project, agile risk planning and people management (taking into account team motivation, personality of members and team balance). Teams aim to maintain a cohesive self-organizing group with established communication protocols where professionalism and ethics are clearly central to team behavior, and people are valued over processes and tools as members work through sprints and releases.
- *Coding Practice*–The team implements rigorous coding practices, including automated testing, developing readable code (with comments), good configuration management (VM, system building, change management, release management) practices, implement standardize headers (for features, modules and components), and employ good peer review. The team's codebase is shared and clearly linked to user stories in the backlog and issues in the issue tracker, where development is done incrementally.
- *Quality Management*–Software quality management processes are adopted to ensure that the software meets clients' specification (user stories). Good software testing is performed, software quality attributes are evaluated (e.g., security, reliability, usability) against established standards, thorough testing is performed before check-in, peer reviews are performed, frequent incremental releases are provided to the client and end users and their feedback accommodated, software measurements are considered (e.g., effort to complete sprint against bugs reported thereafter) in helping to promote software quality improvements.

**2) Product Dimensions (measured via evaluators' scores):**
- *Functionality Scope*–Adequate functionality provided in alignment with requirements promises at the time of project scoping and as the project user stories evolve.
- *Software Quality*–The software solution is of a high quality, is robust and usable. Usability here considers ease of use, look and feel, learnability, efficiency, memorability, errors, and users' overall satisfaction when using the system.
- *Software Design*–The software is adequately designed, with suitable architecture and code (being up to date). There should be clear evidence that the final software architecture has evolved to a stable state and the code should be of high quality in terms of supporting future software maintenance.
- *Software Packaging*–The software is appropriately packaged, with adequate help and documentation. The idea here is that it should be possible to deploy and use the software with the packaged instruction and support. This dimension also covers end user and technical documentation.

---
[2] https://tinyurl.com/3w6smht6

*C. Data Analysis*

We used multiple units of analysis in the study (the project and student). We used statistical and inductive analyses to answer our research questions (RQ1 to RQ3). Demographic data introduced above were used for understanding the projects' context and further explaining the measures. We gaged project effort based on the average number of hours students committed each week. Practice and product measures introduced in the previous section were used for studying practice and product quality. As noted above, students' performances were assessed along these dimensions out of 100 (awarded by a panel of lecturers). We explored variations among projects in answering RQ1. We then analyzed the relationships among variables and modelled practice and product measures to answer RQ2. All measures (including demographic details) were used in our prediction model.

TABLE III. SUMMARY RESEARCH QUESTIONS AND DATA SOURCES FOR DERIVING VARIABLES

| Research Question | Data Source(s) for Deriving Variables |
|---|---|
| RQ1. How do software practice and product performances vary among agile software development teams? | Practice and product measures - evaluated and assigned by lecturers out of 100 based on adherence to processes (guidelines). |
| RQ2. What are the relationships between variations in the use of software development practices and software product performance? | Demographics variables (students' responses), practice and product measures. Practice and product measures evaluated and assigned by lecturers out of 100. |
| RQ3. Are there particular team challenges that affect software practice and product performances during agile software development? | Students' open responses to challenges that affect software practice and product performances (codes extracted through inductive analysis). |

We then adopted an inductive (bottom-up) approach to content analysis to study themes in qualitative data/open ended responses about the challenges that affect software practice and product performances [44]. This analyses took on a qualitative (inductive) tone, where outcomes in response to this question were interpreted in relation to the SEEK knowledge base and earlier themes reported [16]. Outcomes were used to triangulate the earlier quantitative findings. The procedure involved open coding where students' responses were read and re-read for familiarization and initial codes were identified based on explicit, surface level semantics in the data, rather than implicit responses and preconceptions (see Braun and Clarke [45]). Through axial coding, codes were recombined, and connections were formed between ideas. Then, we used thematic mapping to restructure specific codes into broader themes. Finally, following Braun and Clarke's [45] selective coding procedure, the resulting themes were refined and organized into a coherent, internally consistent account, and a narrative ("story") was developed to accompany each theme to answer RQ3. Additional modelling was done with both quantitative and qualitative data in further triangulating our results for the three research questions. Table III provides a summary of the research questions and data sources for deriving the variables.

IV. RESULTS

*A. Demographics*

As noted in the previous section, 85 students agreed to participate in the study. These students had an average age of 21.8 years, and their average year of study was 3.8 years. Of the total students, 15 were enrolled in a Computer Science (CS) degree, 56 in Information Science (IS) and 14 in Software Engineering (SE). Of note is that students of these majors all take software engineering courses, largely leading to software practitioner roles when they graduate. This is reflected in Fig. 1, where most students aspired to work as software developers/engineers or business analysts. Of the total 85 students, 78 were previously employed in a salaried job prior to taking the software engineering course, with 24 of these occupying a software development-related role. While the course the students were enrolled in educated them on software engineering knowledge and principles, previous software project management knowledge was claimed by 66 students, where 59 of these students occupied some role (at work or university) along the software engineering knowledge areas (e.g., analysis and design, testing). Students committed 8.5 hours on average each week on their projects outside the classroom (4-5 hours of work was done each week in the classroom) and the average team size for the teams studied was 5.3 (refer to Table II).

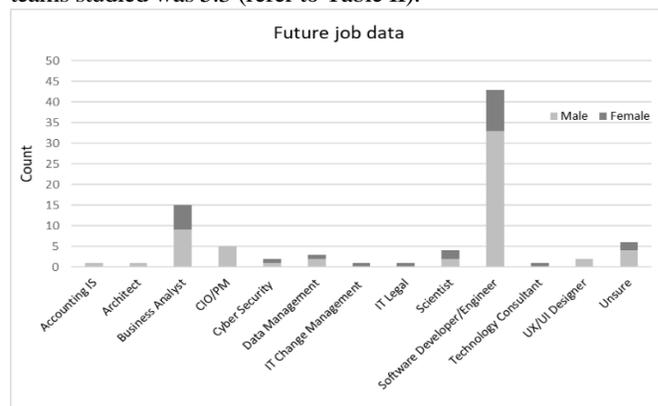

**Fig. 1**. Future job data.

*B. Practice and Product Performances (RQ1)*

We explore variations in software practice and product performances among agile software development teams to answer RQ1. Table IV provides the average scores for the 22 teams along practice and product dimensions, where we can see variations among the teams. Some teams have scored highly in the practice measures but scored lower for the product measures (e.g., System 1 and System 2), while the trend is reversed for other teams (e.g., System 7 and System 18), and yet others scored more consistently across both sets of measures (e.g., System 20 and System 5). We compute an average score for the five practice and four product dimensions and visualize these measures in Fig. 2, where we could see that for most projects the teams' practices were of higher quality than the product quality. However, there were a few exceptions (i.e., for Systems 7, 9 and 18). Table II shows that these latter projects were mobile and web-based, where their team sizes were not outliers. We further examine these patterns in the next section.

*C. Variables Relationships and Predictors of Product Quality (RQ2)*

We formally explore the relationships between variations in the use of software development practices and product quality to answer RQ2. In mitigating for significance by chance, we conducted the more conservative Spearman's correlation test to explore relationships, where the results are provided in Table V (medium to strong correlations are underlined). Here it is seen that older students were typically in a later year of study (to be expected) and they scored lower for the software packaging product measure. Also, those that reported previous knowledge typically occupied a previous software engineering role. We observed that many of the practice quality scores were related, meaning that if teams scored high in one measure they tended to also score high in some of the other practice quality measures. The same was seen for the product measures, where high scores achieved in one product quality measure tended to be linked to high achievement in others. Most noteworthy, and more granular, Table V shows that higher scores achieved for requirements scoping and team planning and communication practices were correlated with higher product design and packaging scores. Higher quality of estimation and scheduling also led to higher design, and higher coding practice and quality management were correlated with all the dimensions of product quality (Functionality Scope, Software Quality, Software Design and Software Packaging). These latter two practice dimensions had the largest effect on software product quality (all dimensions).

TABLE IV. AVERAGE PRACTICE AND PRODUCT QUALITY SCORES FOR TEAMS

| Project Name | Practice Quality | | | | | Product Quality | | | |
|---|---|---|---|---|---|---|---|---|---|
| | Req-Scop | Est-Sch | Plan-Comm | Cod-Prac | Qual-Mng | Func-Scop | Soft-Qual | Soft-Dsg | Soft-Pkg |
| System 1 | 100 | 100 | 100 | 100 | 100 | 86 | 84 | 90 | 86 |
| System 2 | 88 | 88 | 88 | 75 | 70 | 75 | 65 | 75 | 70 |
| System 3 | 87.5 | 95 | 82.5 | 90 | 90 | 82.5 | 80 | 80 | 89.5 |
| System 4 | 83 | 83 | 83 | 81 | 75 | 85 | 75 | 70 | 75 |
| System 5 | 92.5 | 98.8 | 95 | 96.3 | 97.5 | 88 | 87 | 90.5 | 88.5 |
| System 6 | 95 | 95 | 95 | 85 | 90 | 90 | 85 | 75 | 80 |
| System 7 | 88 | 88 | 88 | 90 | 85 | 90 | 85 | 90 | 90 |
| System 8 | 88.8 | 96.3 | 88.8 | 81.3 | 85 | 80 | 77.5 | 90 | 81.8 |
| System 9 | 83 | 83 | 83 | 86 | 80 | 90 | 85 | 85 | 85 |
| System 10 | 90 | 99.2 | 88.3 | 85 | 83.3 | 83.5 | 79.8 | 90.3 | 85 |
| System 11 | 88.8 | 100 | 80 | 85 | 85 | 84 | 78.75 | 90 | 84.75 |
| System 12 | 91 | 99 | 86 | 86 | 88 | 83 | 81 | 91 | 84.8 |
| System 13 | 91.3 | 95 | 93.8 | 91.3 | 95 | 81.5 | 79.75 | 88.75 | 81.75 |
| System 14 | 79 | 79 | 79 | 75 | 60 | 80 | 55 | 85 | 75 |
| System 15 | 98.3 | 98.3 | 96.7 | 86.7 | 96.7 | 89.3 | 84.7 | 87.7 | 88.3 |
| System 16 | 80 | 100 | 75 | 80 | 70 | 80 | 75 | 85 | 80 |
| System 17 | 89 | 98 | 91 | 92 | 96 | 85.2 | 79 | 92 | 86 |
| System 18 | 88.8 | 96.3 | 85 | 83.8 | 88.8 | 89 | 87.5 | 87 | 87.25 |
| System 19 | 86.3 | 97.5 | 90 | 83.8 | 92.5 | 75 | 70 | 90 | 78.75 |
| System 20 | 97 | 97 | 97 | 87 | 85 | 90 | 85 | 90 | 85 |
| System 21 | 89 | 95 | 89 | 89 | 86 | 86 | 82 | 90 | 85.8 |
| System 22 | 88.3 | 96.7 | 95 | 93.3 | 96.7 | 83.7 | 78 | 88.3 | 83.7 |

Note: Req-Scop = Requirements Scoping, Est-Sch = Estimation and Scheduling, Plan-Comm = Team Planning and Communication, Cod-Prac = Coding Practice, Qual-Mng = Quality Management, Func-Scop = Functionality Scope, Soft-Qual = Software Quality, Soft-Dsg = Software Design, Soft-Pkg = Software Packaging

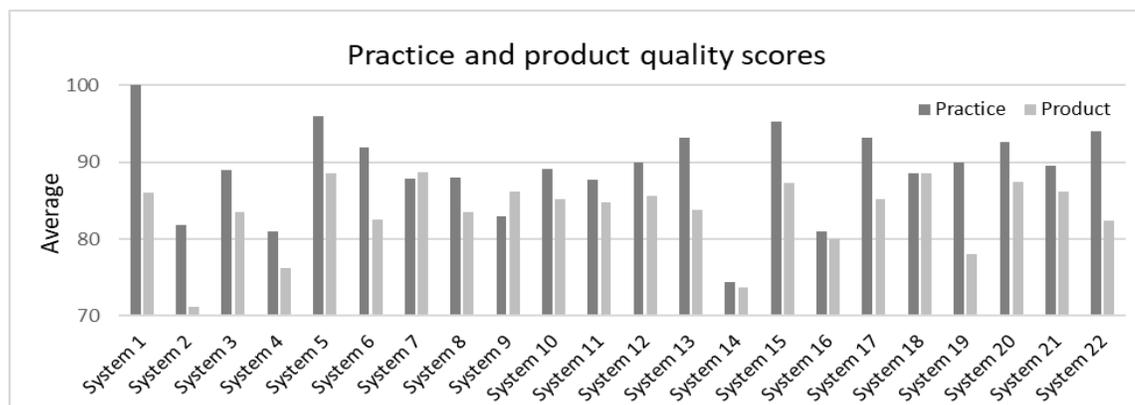

**Fig. 2**. Overall practice and product quality scores for teams.

TABLE V. SPEARMAN'S CORRELATION COEFFICIENTS (RHO) FOR VARIABLES STUDIED

| Variable | 1 | 2 | 3 | 4 | 5 | 6 | 7 | 8 | 9 | 10 | 11 | 12 | 13 | 14 | 15 | 16 | 17 | 18 |
|---|---|---|---|---|---|---|---|---|---|---|---|---|---|---|---|---|---|---|
| 1. Age | 1.000 | -.152 | .412** | .009 | .040 | .014 | -.020 | -.088 | .074 | -.223* | -.222* | -.153 | -.240* | -.147 | -.186 | -.209 | -.287** | -.364** |
| 2. Gender | | 1.000 | .028 | -.193 | -.054 | .007 | -.109 | -.070 | -.009 | .083 | .011 | .067 | .037 | .002 | .087 | .069 | .144 | .005 |
| 3. Study Year | | | 1.000 | -.033 | .155 | .112 | -.080 | -.169 | .099 | .042 | -.201 | .149 | .172 | .099 | .166 | .120 | -.069 | .105 |
| 4. Previous Project | | | | 1.000 | .093 | .298** | .224* | -.025 | -.001 | -.004 | .066 | -.062 | .008 | -.035 | .159 | .194 | -.139 | .116 |
| 5. Previous Job | | | | | 1.000 | .161 | .179 | -.035 | .039 | .003 | -.086 | -.057 | -.013 | -.007 | .054 | .072 | -.006 | .060 |
| 6. Previous Knowledge | | | | | | 1.000 | .401** | .081 | .159 | .007 | -.025 | .011 | .059 | -.050 | .051 | .060 | -.041 | .028 |
| 7. Previous Role | | | | | | | 1.000 | .043 | .057 | .098 | .069 | .055 | .137 | .088 | .216* | .266* | .101 | .208 |
| 8. Hours Committed | | | | | | | | 1.000 | -.132 | -.010 | -.056 | -.162 | -.156 | -.253* | -.079 | -.088 | -.134 | -.084 |
| 9. Team Size | | | | | | | | | 1.000 | -.249* | -.178 | -.194 | -.072 | -.210 | -.081 | -.019 | .035 | -.139 |
| 10. Practice-Requirements Scoping | | | | | | | | | | 1.000 | .367** | .808** | .365** | .457** | .247* | .185 | .595** | .386** |
| 11. Practice- Estimation Scheduling | | | | | | | | | | | 1.000 | .162 | -.038 | .325** | -.066 | .093 | .595** | .292** |
| 12. Practice-Team Planning and Communication | | | | | | | | | | | | 1.000 | .655** | .729** | .235* | .191 | .471** | .381** |
| 13. Practice-Coding Practice | | | | | | | | | | | | | 1.000 | .791** | .560** | .513** | .327** | .652** |
| 14. Practice-Quality Management | | | | | | | | | | | | | | 1.000 | .361** | .343** | .455** | .575** |
| 15. Product-Functionality Scope | | | | | | | | | | | | | | | 1.000 | .913** | .112 | .752** |
| 16. Product-Software Quality | | | | | | | | | | | | | | | | 1.000 | .168 | .753** |
| 17. Product-Software Design | | | | | | | | | | | | | | | | | 1.000 | .418** |
| 18. Product-Software Packaging | | | | | | | | | | | | | | | | | | 1.000 |

Note: * = p <0.05, ** = p<0.01, medium to strong statistically significant correlations underlined

We went one step further and ran a multiple linear regression model to explore the interaction of practice quality variables (predictors) against overall product quality (dependent variable), where the demographics variables are also added to the model as predictors for assessing their mitigation effect. We recognize that our dataset is of limited size for performing regression analysis (i.e., typically, there should be 10 observations for each independent variable, meaning 14 x 10 = 140, as against 85 in our case), which potentially limits the power of our model [46]. Accordingly, our outcomes are conservative at best; we address this threat in Section VI. The results indicate that the model explained 61.2% of the variance ($R^2$=61.2%) and that the model was a significant predictor of software product quality, $F(14,70)$=7.898, p-value <0.001. The variables that significantly predicted software product quality in our regression model were: Requirements Scoping ($\beta$ = .412, p-value < 0.01), Team Planning and Communication ($\beta$ = -.453, p-value < 0.01) and Coding Practice ($\beta$ = .406, p-value < 0.01). Other variables seen in Table V were not significant predictors of software product quality in our regression model, notwithstanding the significant correlations when modelled in isolation (see Table VI for full model details).

Given that our dataset comprised two levels of data (individual and team), to validate the multiple regression outcome we followed up by running a multilevel linear regression model, accounting for measures at the individual (level 1) and team (level 2) levels, where the results mirrored the multiple regression model. At levels 1 and 2, Requirements Scoping (Estimate = .478, Std. Error = .093, p < 0.001 and Estimate = .464, Std. Error = .092, p < 0.001, respectively) Team Planning and Communication (Estimate = -.486, Std. Error = .090, p < 0.001 and Estimate = -.465, Std. Error = .090, p < 0.001, respectively), and Coding Practice (Estimate = .549, Std. Error = .067, p < 0.001 and Estimate = .527, Std. Error = .068, p < 0.001, respectively) were confirmed as significant predictors of software product quality. Although the same assumptions apply to multilevel and regular linear models (e.g., ANOVA, regression), it should be noted that these approaches do not always return the same pattern of outcomes. The data trends in the slopes and intercepts were the same on both individual and team levels in this study.

TABLE VI. REGRESSION COEFFICIENTS RESULTS

| Model | Unstandardized Coefficients | | Standardized Coefficients | t | Sig. |
|---|---|---|---|---|---|
| | B | Std. Error | Beta | | |
| (Constant) | 32.720 | 14.039 | | 2.331 | .023 |
| Age | -.445 | .237 | -.150 | -1.879 | .064 |
| Gender | 1.793 | 1.088 | .128 | 1.648 | .104 |
| Study Year | .616 | .623 | .080 | .989 | .326 |
| Previous Project | 1.760 | 1.903 | .076 | .925 | .358 |
| Previous Job | -.180 | 1.127 | -.013 | -.160 | .873 |
| Previous Knowledge | -.303 | .740 | -.037 | -.409 | .684 |
| Previous Role | 1.162 | .706 | .145 | 1.647 | .104 |
| Hours Committed | -.084 | .159 | -.045 | -.527 | .600 |
| Team Size | 1.444 | .998 | .122 | 1.447 | .152 |
| Requirements Scoping | .412 | .122 | .457 | 3.375 | .001 |
| Estimation Scheduling | .066 | .106 | .072 | .622 | .536 |
| Team Planning and Communication | -.453 | .111 | -.612 | -4.080 | <.001 |
| Coding Practice | .406 | .110 | .581 | 3.678 | <.001 |
| Quality Management | .092 | .104 | .168 | .883 | .380 |

*D. Team Challenges (RQ3)*

We next explored the particular team challenges that affect software practice and product performances during agile software development in answering RQ3. Of the 85 students studied, 52 students provided a response in relation to project challenges. We coded these responses as noted in Section III.C, with a summary provided in Fig. 3. Here it can be seen that lack of technical skills (mentioned by 49 of 52 students) and poor time management (mentioned by 9 students) stood out. Teams particularly singled out new learning required for their technology stack. For instance, a member from System 2 noted "…We faced a difficult learning curve in terms of the technologies we chose. It was not the ideal combination, but we were encouraged to try something new, so we did..." Another member from System 17 alluded to the lack of technical skills being a burden to managing project deadlines and team motivation, noting "…Working with team members who were lacking in technical skills while still ensuring tasks were completed was the biggest challenge and took significant time and energy…" This latter comment highlights the intersection between the lack of technical skills, time management and motivation. Another member from System 13 commented on the challenge of prioritizing project ideas in light of project time constraints, explaining "…Time management was a bit of a struggle, we had so many ideas but struggled to prioritize which ones we would be able to get done over the course of the project..."

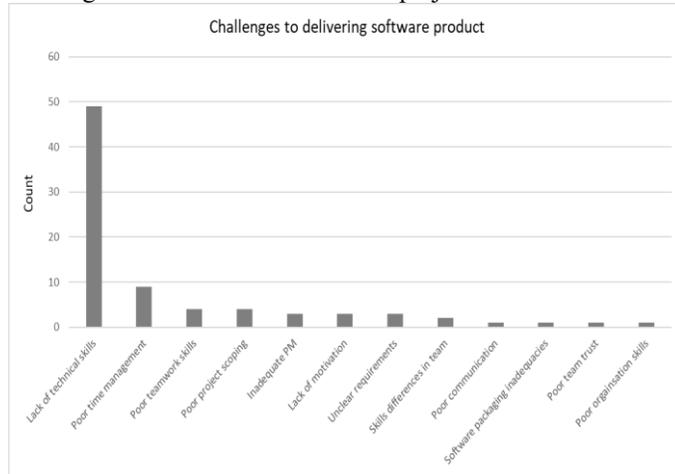

**Fig. 3**. Software development project challenges.

We went one step further to evaluate the effects of these project challenges on project performance. We converted the codes extracted from the coding process to features, before adding them to the other features in Table V as predictors (for the 85 students' dataset). We then ran another multiple linear regression model with all the features as predictors and the overall product quality score as the outcome/dependent variable. While the model was significant ($p < 0.01$), none of the challenges reported from the inductive analysis were statistically significant predictors of software product quality (the most noteworthy challenge was Poor Organization Skills ($\beta = -.545$, p-value = 0.082)). We believe this lack of significance may be due to data scarcity (as seen in Fig. 3 for most themes, apart from "Lack of technical skills"), and so this issue requires further investigation.

## V. DISCUSSION AND IMPLICATIONS

*RQ1. How do software practice and product performances vary among agile software development teams?*

Our findings show that at times there were instances where the final software product was of a higher quality than the practice quality demonstrated during software development. In fact, some teams scored highly for individual practice measures but scored lower for the product measures, while the trend was reversed for other teams, and yet others scored consistently across both sets of measures (practice and product measures).

Outcomes above show that **teams do not all perform the same, even when given the same opportunities**. For instance, all teams implemented their projects using (Scrum) agile practices, involving incremental development with some up-front analysis to understand the domain and decompose initial (but evolving) business requirements. These processes were mentored and supervised, encouraging students' confidence, suggesting that, contrary to earlier evidence reported [8, 10], **agile practices on their own may not lead to the best deliverable**. While teams' maturity was shown to affect their commitment and willingness to address feedback, and hence their performance [21], related variables (e.g., study year, previous project involvement, previous job, previous knowledge) did not impact on teams' product quality in this work.

Students practised weekly stand-ups, weekly retrospectives, team meetings, attended coaching and feedback sessions with knowledgeable staff, and attended weekly technical tutorials on tools and technologies (e.g., build systems and configuration management, version control, unit testing and continuous integration). Students also aimed for frequent software delivery and incorporated continuous client feedback. As part of the software delivery, students were guided on good software packaging and deployment techniques, and received support on the use of software metrics for product and process improvement over the development of their product. Notwithstanding these interventions, which are held to aid productivity [11, 23], students did not perform uniformly.

Findings here suggest that **commitment to processes needs balance**, as notwithstanding these interventions, there were a few instances when **less rigor in process commitment did not lead to poorer overall final software product outcomes**. Also, the optimal process may not be achievable or may not lead to a perfect product, as in one case (System 1), the team conformed fully and exceptionally to all the agile processes during project development but did not deliver a perfect product, in contrast to earlier evidence [28]. For instance, the final software delivered for System 1 was well designed, with suitable architecture and up to date code. There was clear evidence that the final deliverable evolved to a stable state and the code was maintainable. However, the final software solution needed usability improvements.

*RQ2. What are the relationships between variations in the use of software development practices and software product performance?*

Outcomes in this work show that **software development teams that commit to software practice quality seem to do so across the board**. We observed that many of the practice quality scores were related, meaning that if teams scored high in one measure they tended to also score high in some of the other practice quality measures, suggesting that these teams were very committed to following agile practices. Individual software product quality measures were also somewhat related, in that when teams achieved high scores in one product quality measure, they tended to achieve high achievement in others. That said, a team could developed a highly usable piece of software but then fall short on sufficient number of features for the system to be usable.

We found that **higher scores achieved for requirements scoping and team planning and communication practices were correlated with higher product design and packaging scores**.

This outcome support earlier evidence [11, 29], which shows that the quality of requirements impacts on teams' productivity. In the current work, user stories were used as requirements, where teams captured requirements iteratively but were expected to cover the full list of client features, services and constraints. Students were also required to properly express requirements as user stories, and changes were facilitated over the project duration, where the product backlog was used as recommended in the agile body of knowledge [47]. This ensured consistency in the understanding of features and shared context among team members, which could be critical for software design and subsequent system deployment. Team planning covered collaborative, iterative, continuous work on the project, agile risk planning and people management (taking into account team motivation, personality of members and team balance). Teams aimed to maintain a cohesive self-organizing group with established communication protocols where professionalism and ethics were meant to be central to team behavior, and people were valued over processes and tools as members worked through sprints and releases. Not surprising, **teams that scored well for this aspect also delivered adequately designed code of high quality in terms of supporting future software maintenance**.

In fact, **higher quality of estimation and scheduling also led to higher design**. Teams that scored highly for this dimension used planning poker properly, where estimates were realistic and accurate. Scheduling of user stories in sprints took into account feature dependencies, user story size, team members' ability and the team's velocity. The product backlog, release backlog and sprint backlog were used appropriately in aiding scheduling activities, and the team demonstrated a shared vision around the use of these repositories and artefacts. It appears like commitment to such processes helped teams to make sound design decision, leading to good quality software. The shared context around features and their dependencies at the time of estimation seems to help during software design in supporting the team's design decisions, and subsequent software evolution. This relationship has not been reported on previously in the literature.

Finally, **higher coding practice and quality management were positively correlated with all the dimensions of product quality** (Functionality Scope, Software Quality, Software Design and Software Packaging). These latter two practice dimensions (coding practice and quality management) had the largest effect on software product quality (all dimensions). In the consideration of coding practice, teams studied were meant to implement automated testing, develop readable code (with comments), employ good configuration management (version management, system building, change management, release management) practices, implement standardize headers (for features, modules and components), and employ good peer review. The teams' codebases were meant to be shared and clearly linked to user stories in the product backlog and issues in the issue tracker, where development was done incrementally. Software quality management practices focused on software testing, the evaluation of software quality attributes (e.g., security, reliability, usability) against established standards, the conduct of peer reviews, frequent incremental releases for clients' and end users' feedback which is accommodated, and the consideration of software measurements (e.g., effort to complete sprint against bugs reported thereafter) in helping to promote software quality improvements. Not surprising, the consideration of these aspects had the highest effect on software quality. While some of these practices have been singled out previously (e.g., [21] and [9] established that teams that addressed feedback and produce software incrementally produced higher quality products), previous work has not examined the full breadth of processes considered in this study, nor have previous work observed these practices at play.

When examining the overall product quality measures (all measures combined as a single score), our prediction model reveals that **software product quality was predicted by requirements scoping, team planning and communication and coding practice**. Strikingly however, the coefficient for team planning was in the opposite direction (negative). Our outcome here suggests that team planning and communication may not be critical for all aspects of software development. For instance, these practices may be more important for product design and packaging (established in this study) than for developing functionality of appropriate scope and engaging in software quality endeavors. Good coding practice and quality management have a higher influence on the latter dimensions. Agile methods indeed argue for the correct amount of processes and tools [36], and so our outcomes here are understandable.

*RQ3. Are there particular team challenges that affect software practice and product performances during agile software development?*

**Lack of technical skills and poor time management were reported as the biggest challenges that affect software development practice and product performances during agile development**. Student teams particularly singled out the new learning required for their technology stack. Accordingly, there is need to plan for project learning and discovery during project scoping and estimation. Failure to consider time for upskilling members during development, and especially for new technology used for the project may create project pressure. This pressure may result in hasty short-term decisions, leading to the accumulation of technical debt [1, 48]. For instance, the team of System 2 used new technologies that were not familiar to the members to develop their project. During the project development, processes for requirements scoping, estimation and scheduling and team planning and communication were implemented at a very high level. However, once technology challenges were encountered during project implementation/coding, the team's coding and quality management practices slipped away. This then affected all aspects of the final product outcomes, supporting our evidence for coding practice and quality management having the most notable impact on product quality.

In fact, we observed that a **lack of technical skills becomes a significant burden to teams managing project deadlines and team motivation**, especially if teams are new to the technology stack. Deadline pressures and poor motivation of the team in turn may lead to substandard work, and ultimately poor final product outcomes. For instance, the members of System 17 singled out their lack of technical skills being a burden to managing project deadlines and team motivation. However, over their project development the team performed most practices at a very high level. Towards the final software product delivery however, diminished team motivation seems to have affected the team's focus on software quality (e.g., software ease of use and look and feel). While the team delivered a large amount of features and these were well developed and packaged, the user interfaces needed improvement.

Overall, **evidence in this work seems to point to an intersection between the lack of technical skills, time management and motivation**, with the latter two dimensions also interplaying to derail team performance on their own at times. Teams can be challenged prioritizing project ideas in light of project time constraints, sometimes even with the right technical skills. For instance, members of System 13 adopted very good agile practices (and receive high scores) but struggled with deciding on specific project directions, impacting on their final product functionality scope and quality. Agile methods no doubt help with these aspects [29], in terms of bridging challenges among the entire team (via regular reflection sessions) and supporting planning for short iterations and limiting work in progress, where outcomes are assessed frequently [10]. These practices are likely to support team moral, as they have been shown to also benefit team communication and the sharing of tacit knowledge [30]. However, these agile practices can themselves create additional project pressures (reversing empowerment gains [11]), as was seen in this work, leading to poor motivation and threatening project success. **The goal should be to maintain balance, ensuring adequate technical skills for the project and/or time for training and upskilling, good time management, decisive decision making, and in turn, a stimulating low-pressure environment and sustained team motivation.**

## VI. THREATS

While this study provides key insights into the effect of agile practices' quality on software product quality when examining student teams, there are also threats to the reliability and validity of the work. These threats are assessed using established guidelines [49], as follows:

**Reliability**: when considering reliability, survey definition, design, implementation, execution, analysis, and consistency in measures are assessed. The instrument used for collecting data has been assessed for reliability [16] and aligned with software engineering knowledge areas [42]. However, only demographic data and one open question was analyzed from the instrument in this work, reducing reliability threats associated with using solely survey data. Students completed the instrument at the end of the course when all assessments were completed. At this time they agreed to participate in the study and the use of their anonymized data, and so data collection was insulated from major threats. The students also completed the survey immediately after the course was completed, so we believe their reflections and responses benefitted from currency of their knowledge, drawing claims to reliability. We also observed a very good response rate of 70.8%. The University of Otago ethical guidelines ensured that data were collected ethically, thus students' rights were preserved during the data collection process.

The analysis of the demographics data did not involve any interpretation or translation, and thus does not suffer a threat to reliability. Project data and assessment grades analyzed were derived from staff observations, students' repositories (e.g., Git commits and Taiga entries) and course tools (e.g., blackboard). These data reflect the reality of students' project activities and assessment scores, however, staff observations may pose a threat to the work. For instance, staff may expect students to perform well in follow up assessments (e.g., Quality Management) if they were remarkable in earlier assessments (e.g., Requirements Scoping). To mitigate this threat we moderated all assessments, where at least two academic staff members were involved in grading all assessments. The product was graded independently for twice as much as the practice, which may cause students to commit more to the product assessment dimensions, resulting in a threat to the reliability of our outcomes. That said, our pattern of outcomes does not suggest that this was the case, as more teams scored higher on the practice dimensions than the product dimensions. Also, as with all types of data analysis, the qualitative data analysis suffers from a reliability threat. Thematic analyses of open data inherently introduce threats related to subjectivity and biases of the researchers. However, the initial use of themes revealed by earlier work for triangulation [16] where stringent reliability testing were performed, and the consideration of the SEEK knowledge base [38], were employed to mitigate such threats. In addition, we compared students' comments with their repository artefacts, where we saw convergence (refer to Section V). Further, given that we conducted both quantitative and qualitative analyses, we anticipate that our outcomes benefited from triangulation and rigor.

Each survey response was vetted, where students identified their project details, which were verified. The survey data provided very little scope for students to respond favorably about their colleagues, as such data was not analyzed in this work. Comments about project challenges focused on the project, and demographics details were largely about students' personal data and their previous experience. The main data analyzed were captured during project evolution, with limited opportunity for manipulation by respondents. That said, we concede that our small sample likely limited the power of our models (both regular and multilevel regression), thus, our prediction outcomes are conservative, and require follow up research.

**Validity**: this dimension assesses internal, construct and external validity and consistency. For the instrument used previously [16], measures were defined and presented to the students as part of the instrument, and the instrument was discussed in a session at the end of the course, enhancing the likelihood of internal and construct validity. However, this study only explored the demographic data and students' responses about their project challenges. Given the evaluation of the survey relevance previously and analysis of specific data [16], this also strengthen our claims to an internally and constructively valid instrument. That said, we concede that our data represent the responses of students from one university (University of Otago), where students worked around 13.5 hours each week for 13 weeks in the software engineering course, which affects our claim to external validity. Notwithstanding this assessment, some challenges reported align with those reported for other student groups across three other universities [16], and thus, we believe that our outcomes may generalize to projects where students use agile practices (e.g., incremental development, unit testing and continuous integration). However, notwithstanding that 59 of the students studied occupied some role (at work or university) along the software engineering knowledge areas, our outcomes may not generalize to industry settings, where practitioners are likely to be much more experienced software engineers.

## VII. SUMMARY AND FUTURE WORK

Agile methods and related practices have been touted for the value they provide to the software development process, with research evidence supporting this position. However, much of this evidence is provided using surveys and interview studies, where practitioners tend to provide their reflections. Limited research observe and measure agile practices and link these observations to software product outcomes, due to challenges associated with gaining access to practitioners' data. Students' projects have thus aided our understanding of agile teams. We contribute to such endeavors in this work, and examined the software artefacts and development practices of 22 student teams comprising 85 students altogether. We conducted a survey where we gathered various demographics details and comments in relation to project challenges. These responses were combined with several practice and product measures to explore how software practice and product performances vary among agile software development teams, what are the relationships between variations in the use of software development practices and software product performance, and the team challenges that affect software practice and product performances during agile software development.

Our outcomes show that teams did not perform uniformly, but most often teams had higher practice quality than product quality. That said, there were exceptions where products delivered demonstrated higher quality than the practices that were used to develop them. This points to the need for balancing the use of processes, where team should be cautious that there is no perfect or optimum practice configuration for delivering successful software. Findings show that teams tended to score highly across both practice and product quality dimensions whenever they performed well. Also, while some practice quality dimensions were linked to product quality dimensions (e.g., higher scores achieved for requirements scoping and team planning and communication practices were correlated with higher product design and packaging scores), our most significant finding was that higher coding practice and quality management were positively correlated with all dimensions of product quality (Functionality Scope, Software Quality, Software Design and Software Packaging).

Our prediction model reveals that software product quality was predicted by requirements scoping, team planning and communication, and coding practice. However, high levels of team planning and communication were not necessary for all software development activities. When examining project challenges, we observed the lack of technical skills and poor time management to be the most frequently mentioned challenges. Evidence in this work also points to an intersection between the lack of technical skills, time management and motivation, with the latter two dimensions also interplaying to derail team performance on their own at times. While agile practices may help to mitigate these issues, our evidence suggests that this may also hinder teams. Thus, agile teams assembled to develop software should aim to maintain balance, ensuring adequate technical skills for the project and/or time for training and upskilling, good time management, decisive decision making, and in turn, a stimulating low-pressure environment and sustained team motivation.

We recommend further investigations into agile practice and product outcomes, starting with a larger sample of projects. Studies should also complement our project analysis with student feedback on perceptions on their performance over each practice dimension. Further, research should assess team members' profiles, to explore how team dynamics may potentially mediate our outcomes. Finally, we encourage replication studies.